\documentclass[aps,prl,twocolumn,superscriptaddress,showpacs,floatfix]{revtex4}

\usepackage{graphicx}

\begin{document}

\title{Core percolation and onset of complexity in Boolean
networks}
  
\author{L. Correale} 
\affiliation{Politecnico di Torino, Corso Duca degli Abruzzi 24,
I-10129 Torino, Italy}
\affiliation{ISI Foundation, Viale Settimio Severo 65, Villa Gualino,
I-10133 Torino, Italy}

\author{M. Leone}
\affiliation{ISI Foundation, Viale Settimio Severo 65, Villa Gualino,
I-10133 Torino, Italy}

\author{A. Pagnani}
\affiliation{ISI Foundation, Viale Settimio Severo 65, Villa Gualino,
I-10133 Torino, Italy}

\author{M. Weigt}
\affiliation{ISI Foundation, Viale Settimio Severo 65, Villa Gualino,
I-10133 Torino, Italy}

\author{R. Zecchina}
\affiliation{International Centre for Theoretical Physics, Strada
Costiera 11, P.O. Box 586, I-34100 Trieste, Italy}

\date{\today}

\begin{abstract}
The determination and classification of fixed points of large Boolean
networks is addressed in terms of constraint satisfaction problem. We
develop a general simplification scheme that, removing all those
variables and functions belonging to trivial logical cascades, returns
the computational core of the network. The onset of an easy-to-complex
regulatory phase is introduced as a function of the parameters of the
model, identifying both theoretically and algorithmically the relevant
regulatory variables.
\end{abstract}


\maketitle

{\it Introduction ---} Boolean Networks (BN) are dynamical models
originally introduced by S. Kauffman in the late 60s
\cite{Kauf69}. Since Kauffman's seminal work, BN have been used as
abstract modeling schemes in many different fields, including cell
differentiation, immune response, evolution and gene-regulatory
networks (for an introductory review see \cite{GCK} and references
therein). In recent days, BN have received renewed attention as a
powerful scheme of data analysis and modeling of high-throughput
genome and proteome experiments \cite{IlyaBook2005}.

The central issue of previous research was the description and
classification of different BN attractor types under
deterministic parallel update dynamics
\cite{Kauf69,Kauf1,samuelsson}. Special attention was dedicated to
so-called critical BN \cite{Derrida}, situated at the transition
between ordered and chaotic dynamics. We follow a complementary
approach: Our main goal here is to study the statistical properties of
{\em fixed points} of the dynamics of general BN. We first reformulate
the original dynamical problem into a {\em constraint satisfaction
problem}, and then we study it with techniques borrowed from statistical
mechanics of disordered systems, cf. \cite{martinbook}. As a result, we
are able to classify different types of fixed-point organization, and
to identify the set of relevant regulatory variables.

{\it The model ---} In Kauffman networks, the expression of one gene
is modeled by a Boolean variable: the expression level $x_i$ of a gene
$i$ takes only the values 0 ($i$ is not expressed) or 1 ($i$
is expressed) \cite{Kauf1}. Having $N$ genes, the global expression
pattern of a cell is thus given by a $N$-dimensional vector $\vec
x = (x_1,...,x_N) \in \{0,1\}^N$ with binary entries. Even if this
binary idealization seems to be oversimplified with respect to
biological reality (including mRNA and protein {\it concentrations},
post-transcriptional and post-translational modifications, etc.), it
reflects the frequently only qualitative nature of biological
knowledge. Moreover, it is expected that robust genome-wide biological
processes can be qualitatively understood on this level of
modeling. In this context, one can hope to describe gene regulation
via Boolean functions~\cite{Hwa}: the expression level of a regulated
gene $a$ is determined by the expression levels of the transcription
factors $a_1,...,a_K$:
\begin{equation}
\label{eq:function}
x_a = F_a(x_{a_1},...,x_{a_K})
\end{equation}
with $a\in A\subset \{1,...,N\}$ running over all regulated genes.

\begin{figure}[htb]
\vspace{0.2cm}
\begin{center}
\includegraphics[width=0.62\columnwidth]{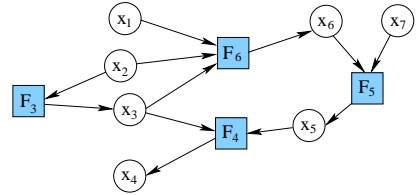}
\end{center}
\caption{Example for a small Boolean network, circles symbolize
variables, squares Boolean functions. $x_1$ is an example for an
external input variable, $x_4$ for a functional variable, whereas
$x_3$ stands for a regulatory variable, see below.}
\label{fig:fig1}
\end{figure}

The central question of this work is: are methods from the statistical
mechanics of disordered systems able to capture number ${\cal N}_{fp}$
and organization of {\it stationary points} of this network, {\em
i.e.}~vectors $\vec x$ fulfilling simultaneously {\it all} equations
of type (\ref{eq:function}) with $a\in A$? The main step here is to
write them as {\it zero-energy ground states} of a cost function, or
Hamiltonian, defined as
\begin{equation}
\label{eq:H}
{\cal H}(\vec x) = \sum_{a\in A} x_a \oplus 
F_a(x_{a_1},...,x_{a_K})\ .
\end{equation}
The symbol $\oplus$ stands for the logical XOR operation, {\em
i.e.}~each cost term contributes zero to the sum iff
Eq.~(\ref{eq:function}) is fulfilled, and one otherwise. The
Hamiltonian ${\cal H}(\vec x)$ thus counts the {\it number of
unsatisfied Boolean equations}. This embeds the problem of finding
fixed points into the class of {\it constraint-satisfaction problems},
which have been recently studied extensively from the point of view of
statistical physics \cite{martinbook}, using various techniques from
spin-glass physics, in particular the cavity method \cite{MPZ}. Note
that a related approach based on topological properties of the BN was
independently proposed in \cite{Lagomarsino}.

In this work we concentrate on the interplay of different types of
functions (represented by $F_a$) for given probability measure of the
topology of the network (represented by the $a,a_1,...,a_K$). For the
moment we restrict the analysis to the case of only $K=2$
inputs. Extensions will be discussed at the end of this Letter.

\begin{figure}[htb]
\vspace{0.2cm}
\begin{center}
\includegraphics[width=0.9\columnwidth]{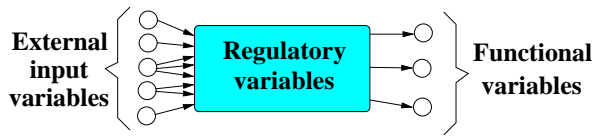}
\end{center}
\caption{{\it External input variables:} There are $N-M$ variables
which are not regulated by any function. They represent external
inputs to the network. {\it Regulatory variables} are all those
variables which are regulated and regulate. {\it Functional
variables:} There are $e^{-2\alpha} N$ variables which do not regulate
any other function.}
\label{sandwich}
\end{figure}

Let us assume that the network is random, {\em i.e.}~its topology is
described by a random directed hyper-graph where the triples
$(a,a_1,a_2)$ are randomly chosen with the following conditions: 
{\it (i)} A function is regulated by at most one
function, {\em i.e.}~any two distinct triples $(a,a_1,a_2)$ and
$(b,b_1,b_2)$ fulfill automatically $a\neq b$. {\it (ii)} There are
$M=\alpha N$ of these triples. Condition {\it (i)} restricts
the fraction $\alpha =M/N$ to the interval $[0,1]$. We can distinguish 
three relevant types of variables as displayed in 
Fig.~\ref{sandwich}.

We now have to specify the functions acting on top of the random
topology.  There are $2^{2^K}=16$ Boolean functions, which can be
grouped into 4 classes \cite{Kauf2}: (i) The two constant
functions. (ii) Four functions depending only on one of the two
inputs, {\em i.e.} $x_1,\overline x_1, x_2,\overline x_2$. (iii) {\it
AND-OR class:} There are eight functions, which are given by the
logical AND or OR of the two input variables, or of their
negations. These functions are {\it canalizing} in the sense that,
e.g., if $F(x_1,x_2)=x_1 \wedge x_2$ and the value of $x_1$ is set to
zero, the output is fixed to zero independently of the value of
$x_2$. It is said that $x_1$ is a {\it canalizing variable} of $F$
with the {\it canalizing value} zero. (iv) {\it XOR class:} The last
two functions are the XOR of the inputs, and its negation. They are
not canalizing: Whatever input is changed, the output changes, too.

Here we are only interested in true $K=2$ functions, {\em i.e.}~in the
AND-OR class and the XOR class, but the extension of our analysis to
the whole spectrum of $K=2$-functions is a minor technical point. The
organization of fixed points does not depend on the relative
appearance of different functions within each class, but only on the
relative appearance of classes. We therefore require $xM$ functions to
be in the XOR class, and the remaining $(1-x)M$ functions to be of
AND-OR type, with $0\leq x\leq 1$ being a free model parameter. In
this sense, for $K=2$, the network ensemble is completely defined by
$\alpha$ and $x$.

{\it Computational core ---} Many variables can be fixed following
logical cascades starting in external variables, and the corresponding
Boolean equations become fulfilled. Other functions can be satisfied
trivially because they include variables being otherwise
unconstrained (see below). However, a certain set of equations
(depending both on the BN and the configuration of external
variables), cannot be satisfied on the basis of such simple local
arguments. We define the {\it computational core} (CC) as the maximal
subnetwork formed by these functions.

The first process to be considered in this context is {\it propagation
of external regulation} (PER). If both inputs of a function are fixed
externally, or if a canalizing input is fixed to its canalizing value,
also the output is directly fixed externally. This process can be
propagated, exploiting all direct logical implications of the
configuration of the external input variables: all the functions
which are trivially satisfied by this process can be removed. The
probability that a function survives is
$$
\pi_{PER} = 1-(1-\alpha \pi_{PER})^2
- (1-x) (1-\alpha \pi_{PER}) \alpha \pi_{PER}\ ,
$$ {\em i.e.}~neither both input variables are fixed by PER nor, in
the case of a canalizing function, a canalizing variable is fixed by
PER to its canalizing value. For small $\alpha$, this equation has
only the trivial solution $\pi_{PER}=0$, {\em i.e.}~all variables are
completely fixed by PER with probability one. At $\alpha_{PER} (x) =
1/(1+x)$, a new solution appears continuously. Above this point, a
finite fraction of all functions survives the PER decimation process,
forming the PER core.
The core percolation point moves toward one in the limit $x\to 0$: in
a pure AND-OR network an extensive PER core never exists due to the
canalizing character of the functions. PER is closely related to the
{\em percolation of information} discussed in \cite{GCK}, which
studies the parallel evolution of two slightly distinct
configurations. Identifying external variables in our model with
constant functions in Kauffman's original formulation, the phase
transition line becomes exactly the location of critical BN. The
region left to the PER transition line corresponds to ordered, while the
region on its right corresponds to chaotic networks under parallel dynamics 
\cite{more}.

\begin{figure}[htb]
\vspace{0.3cm}
\begin{center}
\includegraphics[width=0.9\columnwidth]{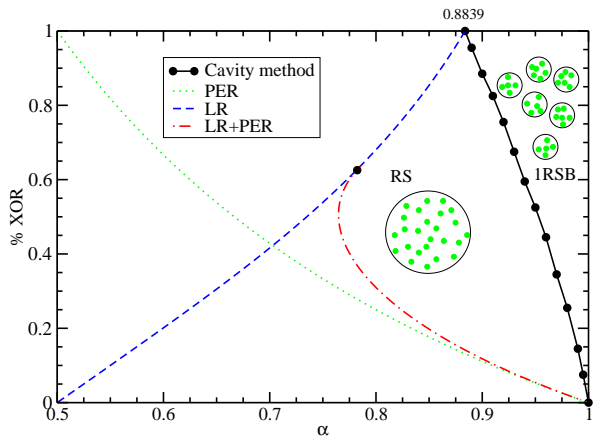}
\end{center}
\caption{Phase diagram for $K=2$: The dotted line gives the
percolation transition for the PER core, the dashed one for the LR
core. The dash-dotted curve results from the combination of LR and
PER, and merges in the tricritical point (0.785,0.636) with the LR
curve. Below this point, the transition is continuous, above
discontinuous in the core size. The full line indicates the RSB
transition from an unstructured solution set to a highly clustered
one.}
\label{fig:phasediagram}
\end{figure}

The second process in determining the core is the {\it leaf-removal
procedure} (LR). A leaf is a variable which has degree one, {\em i.e.}
a variable which is either functional (regulated but not regulating)
or which is not regulated, and acts exactly on one other variable. A
regulated leaf corresponds to a Boolean equation which can be
trivially fulfilled, the same is true if both input variables of a
(non-constant) function are leaves. In the XOR case, even a simple
input leaf allows for satisfying the Boolean equation independently of
the values of the other input and the output \cite{XOR}. None of these
cases induces complex behavior based on frustration - meaning in this
context that the fixed points of the Boolean network are easily
determined - and can be eliminated from the network. This process may
induce new leaves, and can be iterated until no removable function is
left. The survival probability of a randomly selected Boolean function
reads now:
$$
\pi_{LR} = x (1-t_{out})(1-t_{in})^2 + (1-x)(1-t_{out}) (1-t_{in}^2)
$$ 
It depends still on the probability that a variable at a certain point
in this process becomes a regulated ($t_{out}$) or regulating
($t_{in}$) leaf, which can be determined self-consistently using an
iterative argument:
\begin{eqnarray}
t_{out} &=& \exp\{ -2\alpha(1-t_{out})(1-x t_{in}) \}
\nonumber\\
t_{in} &=& \left( 1-\alpha x (1-t_{in})^2-\alpha (1-x) (1-t_{in}^2)
\right) t_{out}\ .
\nonumber
\end{eqnarray}
There is always the solution $\pi_{LR} = 0$ corresponding to complete
removal of all functions by LR. At some $x$-dependent
$\alpha_{LR}(x)$, a second solution appears discontinuously, and the
LR core size jumps to a finite fraction of the complete network. The line
$\alpha_{LR}(x)$ is monotonously growing in $x$, going from 1/2 for
the pure AND-OR case to 0.8839 for the pure XOR case.

Both procedures can be combined in order to determine the {\it
computational core} of the Boolean network under one external
condition: First LR is applied, and then PER is needed to see which
variables on the LR core are fixed via propagation of external
regulation. The problem can still be solved analytically. Here we give
only the results, technical details will be presented elsewhere
\cite{more}. For $x>0.636$, the emergence of the CC is discontinuous
and appears exactly at the point $\alpha_{LR}(x)$. The LR core thus
contains immediately an extensive computational core. The height of
the discontinuity decreases, however, if we approach $x\simeq 0.636$
from above, and we find a tricritical point in
$(x,\alpha)=(0.636,0,785)$. Below this point, the CC emerges only
after the LR percolation transition, and the transition is
continuous. The results are summarized in the phase diagram of
Fig.~\ref{fig:phasediagram}.

\begin{figure}[htb]
\vspace{0.2cm}
\begin{center}
\includegraphics[width=0.9\columnwidth]{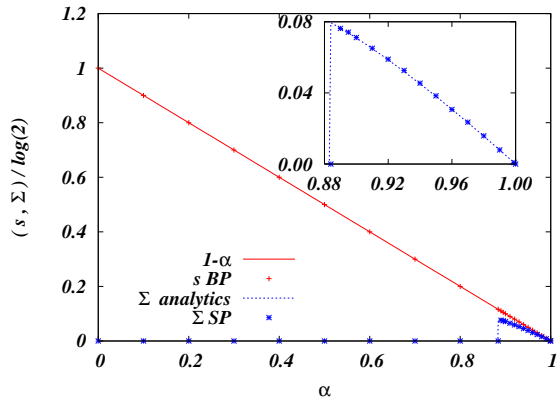}
\end{center}
\caption{Entropy $s$ and complexity $\Sigma$ as a function of
$\alpha$, here for $x=1$. We display both the analytical result and
single sample values using belief and survey propagation (BP,SP)
\cite{MPZ,more}. The complexity curves for $x<1$ are similar, but the
discontinuous onset of the clustered phase is shifted to higher
$\alpha_d(x)$.}
\label{fig:entropy}
\end{figure}

{\it Clustering of solutions and complexity ---} What does the
existence of an extensive CC imply in the structure and organization
of fixed points? This questions can be answered by applying the {\it
zero-temperature cavity method} of statistical mechanics to the
Hamiltonian ${\cal H}(\vec x)$ defined in Eq.~(\ref{eq:H}); technical
details will be presented in a longer publication \cite{more}.
The first result concerns the {\it solution entropy}
which is found to be
$$
s = \frac 1N \overline{ \log( {\cal N}_{fp}) } = (1-\alpha) \log(2)
$$ 
with the over-bar denoting the average over the random network
ensemble with fixed $\alpha$ and $x$ (with, in fact, no dependence on
$x$). This implies that fixed points satisfying all Boolean functions
exist for all $\alpha<1$ and all $x$. The line $\alpha=1$ can be
considered as the SAT/UNSAT transition line. The entropy value shows
that each configuration of the $(1-\alpha)N$ external input variables
leads on average to one stationary point, {\em i.e.}  the state of all
internal and functional variables is mainly determined by the external
conditions, even in the case of a PER core, where this fixing is not
just a simple logical implication.
Even if the entropy is analytic in all the phase diagram, the
organization of the fixed points in configuration space undergoes a
clustering transition at some $\alpha_d(x)$, see
Fig.~\ref{fig:phasediagram}. Below it, all solutions are
collected in one cluster: any pair of them can be connected by a
path via other solutions, where in each step only a finite number of
variables can be changed. The solution space is technically
named replica symmetric (RS). At the clustering transition, the replica 
symmetry breaks
spontaneously (RSB): an exponential number of macroscopically
separated clusters of fixed points appears. Their number, or more
precisely its normalized logarithm, is called {\it complexity}. It
jumps discontinuously at $\alpha_d(x)$, as can be seen in
Fig.~\ref{fig:entropy}.
The clustered phase sets on in general well inside the
region where a CC exist. Exceptions are the
extreme points $x=0$ (neither core nor clustering appear at any
$\alpha<1$) and $x=1$ (both transition points coincide).

{\it Beyond $K=2$ ---} The case of Boolean functions depending on
exactly $K=2$ can be generalized. However, the number of Boolean
functions of $K$ variables diverges as $2^{2^K}$. For $K=3$, the 256
functions can still be classified completely: there are 14 classes, 4
of them are the classes already discussed in the context of $K=2$, and
only 10 lead to actual $K=3$ functions. For $K\geq 4$, 
non-exhaustive numerical checks were done. The main results are: (i) The
minimum of the clustering point $\alpha_d$ over all combinations of
functional classes is always given by the pure $K$-XOR class (the only
completely analytically accessible). We find $\alpha_d=0.782,\ 0.699$
for $K=3,\ 4$ and $\alpha_d=(\ln K + \ln\ln K + \ln\ln\ln K+...)/K$
for $K\gg 1$. Note that the range of the complex phase becomes larger
with $K$.  (ii) We conjecture that, for fixed $K$, the only class not
leading to clustering is the pure $K$-AND-OR one. This has been
checked explicitly for $K=3$, and non-exhaustively also for
$K=4$. This class becomes, however, both combinatorially and
biologically less important for higher $K$, where threshold-like
functions are expected to be more relevant.

{\it Conclusion and outlook ---} We have analyzed the
organization of fixed points in random Boolean networks
identifying the sudden emergence of a computational core, whose
existence is a necessary (but not sufficient) condition for a globally
complex phase where all fixed points are organized in an exponential
number of macroscopically separated clusters. This phenomenon is found
to be robust with respect to the choice of the Boolean functions, and
missing only in networks completely made of AND and OR functions. In
addition, the size of the complex regulatory phase grows if a higher
number $K$ of inputs to the Boolean functions is allowed.

{\it Open Questions ---} Our analysis is based on ground states of a
Hamiltonian counting the number of violated functions, not
directly related to any overlying biologically motivated dynamics. 
The {\it dynamical accessibility} of fixed points remains
an open challenge which will be addressed in a future
project. Moreover, noise sources present in real
cases may force the network to quasi-stationary points close to the
fixed ones (some regulation might not always function properly,
without effecting the overall state of a cell). In this view, the
study of the organization and accessibility of meta-stable states in
the region of low complexity and in the case of fixed external inputs,
together with their relation with quasi-stationary points might be
very relevant.

A second interesting issue is moving from average case analysis to
real networks samples. Our method can be directly implemented as a
{\it message passing algorithm on single network instances}, and not
only on random ensembles as discussed in this work. However, very few
genome-wide data are so far available. In particular, for
multi-cellular organisms only small functional modules for
well-described functions are known, and large-scale networks known for
yeast \cite{GuBoBoKe} and {\it E. coli} \cite{ShMiAl} contain only the
topological structure, not an extensive description of the regulatory
functions. It is thus highly interesting to infer gene regulation
networks from experimentally easily accessible high-throughput
experiments, as e.g.~given by genome-wide expression profiles
\cite{600Genes}. This inverse problem could be treated with tools
similar to the ones used in the present analysis.

{\it Acknowledgment ---} We deeply thank Alfredo Braunstein for
allowing us to use his SP implementation. This work was supported by
EVERGROW (integrated project No. 1935, EU 6th Framework Programme).

\end{document}